\documentclass[aps,pra,reprint, amsmath, amssymb,superscriptaddress,nofootinbib]{revtex4-1}

\usepackage{bm}
\usepackage[retainorgcmds]{IEEEtrantools}
\usepackage{graphicx}
\usepackage{mathrsfs}
\usepackage{amsmath}
\usepackage{amssymb}
\usepackage{color}
\usepackage{amsfonts}
\usepackage{times,txfonts}
\usepackage{nicefrac}
\usepackage[colorlinks=true,linkcolor=blue,urlcolor=blue,citecolor=blue,pdfusetitle]{hyperref}

\newcommand{\tr}{\text{tr}}

\begin{document}

\title{Upper bound for quantum entropy production from entropy flux}
\date{\today}
\author{Domingos Salazar}
\affiliation{Unidade de Educa\c c\~ao a Dist\^ancia e Tecnologia,
Universidade Federal Rural de Pernambuco,
52171-900 Recife, Pernambuco, Brazil}

\begin{abstract}

Entropy production is a key quantity characterizing nonequilibrium systems. 
However, it can often be difficult to compute in practice, as it requires detailed information about the system and the dynamics it undergoes. 
This becomes even more difficult in the quantum domain, and if one is interested in generic nonequilibrium reservoirs, for which the standard thermal recipes no longer apply.
In this paper, we derive an upper bound for the entropy production in terms of the entropy flux for a class of systems for which the flux is given in terms of a system's observable. 
Since currents are often easily accessible in this class of systems, this bound should prove useful for estimating the entropy production in a broad variety of processes. 
We illustrate the applicability of the bound by considering a three-level maser engine and a system interacting with a squeezed bath. 

\end{abstract}
\maketitle{}

{\bf \emph{Introduction -}} The degree of irreversibility of nonequilibrium processes can be characterized by the entropy production, the key quantity appearing in the second law of thermodynamics. 
While already present in the seminal papers by Clausius~\cite{Clausius1854,Clausius1865}, this concept gains additional significance when thermodynamics is extended to the stochastic and quantum regimes. Indeed, a plethora of different processes, from work extraction, to heat exchange, can all be described in terms of entropy production. 

Moreover, the development of fluctuation theorems~\cite{Evans1977,Gallavotti1995,Crooks1998,Jarzynskia2008,Jarzynski1997,Jarzynski2000,Crooks1998,Hanggi2015,Saito2008}, which can be cast in a unified language in terms of entropy production, further corroborates the usefulness of this concept~\cite{Esposito2009,Campisi2011}. In all cases, the entropy production has a general form,
\begin{equation}
\label{equation1}
    \Sigma = \Delta S + \Phi,
\end{equation}
where $S=-\tr (\rho_S \ln \rho_S)$ is the von Neuman's entropy, $\rho_S$ is the system's density matrix and $\Phi$ is the entropy flux. 

Unfortunately, the entropy production can be notoriously difficult to assess, particularly for quantum systems~\cite{Manzano2018,Landi2020}. 
First, it is not a physical observable~\cite{DeGroot1961}. 
In order to relate it to observables, one must therefore know the specific type of dynamics the system is undergoing (e.g., Langevin equation, master equation, nonequilibrium Green's function, etc.).
For standard thermal processes, one can usually relate the entropy production to the flow of heat. 
But for systems interacting with nonequilibrium reservoirs, this is no longer possible. 
In fact, quite often computing the entropy production requires full knowledge of the reservoirs interacting with the system, as well as the system-bath interactions~\cite{DeChiara2018}. 
It may thus be that assessing the entropy production is possible only via full tomography of both system and environment, something which easily becomes prohibitive. 

In this paper we derive an upper bound for the entropy production, which is cast in terms of the entropy flux to the environments, for a class of systems for which the entropy flux is given in terms of some particular state $\rho^*$, $\Phi=\tr\{(\rho_S(t)-\rho_S(0))\ln\rho^*\}$, for all times, meaning it is written as a system's observable. This is the case, for instance, in situations where the unitary
operator of the system plus environment has a global fixed point, as discussed in the next section.

The initial state of the system $\rho_S(0)$ is known, but the nonequilibrium state $\rho_S(t)$ and the dynamics are not. Our result is derived using a variation of the MaxEnt principle~\cite{Jaynes1957}. 
Recently, this principle has been applied to derive tighter bounds for heat exchange~\cite{Timpanaro2019a,Strasberg2021A}.
Our results are similar in spirit, but address fundamentally different questions. We show that the entropy production has an upper bound in terms of the entropy flux rate,  $\phi(t)=d\Phi/dt$. It reads
 \begin{equation}
\label{upperbound}
    \Sigma \leq \Sigma_a+ \int_0^t [1-\alpha(t)]\phi(t)dt.
\end{equation}
where  $\Sigma_a=\Sigma_a(\rho(0))$ is the adiabatic entropy production (function of the initial state) and $\alpha(t)$ is a dynamic temperature-like scalar defined implicitly as a function of the known entropy current and initial state through the constraint
\begin{equation}
\label{alpha}
f(\alpha)= \tr(\rho(0)\ln \rho^*)+\Phi,
\end{equation}
where $f(\alpha)=(d/d\alpha)\ln Z(\alpha)$ and $\ln Z(\alpha):=\ln\tr((\rho^*)^\alpha)=(1-\alpha)S_{\alpha}(\rho^*)$, $S_\alpha(\rho^*)$ is the quantum Renyi entropy of $\rho^*$. 
Note that the bound (\ref{upperbound}) is computed solely in terms of three ingredients: the entropy current $\phi(t)$ and the states $\rho(0)$ and $\rho^*$. The usefulness of the bound lies in the fact that currents are usually easily accessible for the class of systems under consideration. The initial coherence in $\rho(0)$ with respect to $\rho^*$ is contained in the term $\Sigma_a$, which increases the bound when compared to the classic counterpart, akin to previous results on coherence resource theory \cite{Lostaglio2015,Streltsov2016a,Baumgratz2014,Santos2019}.

To illustrate the usefulness of our bound, we apply it to two interesting problems in quantum thermodynamics: the three-level maser engine of Scovil and Schulz-DuBois~\cite{Scovil1958,Geusic1967,Mitchison2019,Li2017a}, and a system interacting with a squeezed bath~\cite{Manzano2018b,Manzano2016}.

{\bf \emph{Formalism -}} We consider here a system $S$ prepared in an arbitrary state $\rho_S(0)$, possibly infinite dimensional, and a set of reservoirs $E_1, E_2, E_3, \ldots$, prepared in a product state $\rho_{E_1} \otimes \rho_{E_2} \otimes \ldots$. 
We do not assume that the environments are in thermal states.
The joint $SE_1E_2\ldots$ system interacts by means of a generic unitary $U_t$, which may be generated by arbitrary interactions plus potential external drives. 
This therefore leads to the map $\rho(t) = U_t \rho(0) U_t^\dagger$, where $\rho(0) = \rho_S(0) \otimes \rho_{E_1} \otimes \rho_{E_2}\otimes \ldots$. 

Defining the entropy production for this process requires a framework that holds beyond the usual hypotheses in thermodynamics. 
We adopt here the formalism of~\cite{Esposito2010a}, which casts the entropy production in terms of the correlations built up between system and bath, as well as the changes in the bath's state; \textit{viz.}, 
\begin{equation}\label{sigma_def}
    \Sigma(t) = I_{\rho(t)}(S:E) + D\big(\rho_E(t) || \rho_E(0) \big).
\end{equation}
The first term reads $I_{\rho(t)}(S:E) = S(\rho_S(t)) + S(\rho_E(t)) - S(\rho(t))$, where $S(\rho) = - tr\rho \ln \rho$ is the von Neumann entropy, while $\rho_S(t) = \tr_E \rho(t)$ and $\rho_E(t) = \tr_S \rho(t)$ are the reduced states of the system and the baths (here $E$ is a shorthand notation for all environments $E_1E_2\ldots$).
The second term in~\eqref{sigma_def} is the quantum relative entropy (Kullback-Leibler divergence) $D(\rho || \sigma) = \tr(\rho \ln \rho - \rho \ln \sigma)$, between the final and initial states of the bath. 
The definition~\eqref{sigma_def} recovers standard thermodynamic results when the environments are thermal. Moreover, it has a clear operational interpretation (as first put forth in~\cite{Jaynes1957}),
where irreversibility emerges from the assumption that an external agent does not have access to local operations on the environment, or to any system-environment correlations that are created by the unitary $U$~\cite{Manzano2017a}. 

Since the global dynamics is unitary, it follows that $S(\rho(t)) = S(\rho(0)) = S(\rho_S(0)) + S(\rho_{E_1}) + S(\rho_{E_2}) + \ldots$. 
As a consequence, with some rearranging, one may also rewrite~\eqref{sigma_def} as~\cite{Landi2020}
\begin{equation}
\label{Entropy}
    \Sigma = S(\rho_S(t)) - S(\rho_S(0)) + \Phi(t), 
\end{equation}
which is Eq. (\ref{equation1}), where $\Phi(t)$, called the entropy flux, is given by 
\begin{equation}
\label{Flux0}
    \Phi(t) = -\sum\limits_i \tr\Big\{ \big(\rho_{E_i}(t) - \rho_{E_i} \big) \ln \rho_{E_i}\Big\}.
\end{equation}

We see that the entropy flux is written as a sum of individual contributions from each bath, each given by the change in the ``thermodynamic potential'' $\ln \rho_{E_i}$. 
In the following steps, we consider a specific class of systems for which such environment potential also corresponds to a system potential,
\begin{equation}
    \tr \{(\rho_S(t)-\rho(0))\ln \rho^*)\}+\sum_i \tr \{(\rho_{E_i}(t)-\rho_{E_i})\ln \rho_{E_i}\}=0,
\end{equation}
for some constant system's density matrix $\rho^*$. In this case, the entropy flux is written in terms of system related quantities using (\ref{Flux0}),
\begin{equation}
\label{constraint}
\Phi(t)= \tr\{(\rho_S(t)-\rho_S(0)) \ln \rho^*\},
\end{equation}
that acts as a generalized potential.
This is the case, for instance, in situations where the unitary operator of the system plus environment has a global fixed point $\rho^*=\rho^{ss}$, i.e., $U(\rho^* \otimes \rho_E) U^\dagger = \rho^*\otimes\rho_E$ \cite{Landi2020}. In the case of thermal operations, one has $\rho^*=\rho^{ss}=\exp(-\beta H)/\tr(\exp(-\beta H))$ and the entropy flux has the familiar form $\Phi=-\beta \tr\{(\rho_S(t)-\rho_S(0) )H\}$. In the general case, however, $\rho^*$ is not necessarily the fixed point $\rho^{ss}$ as we are going to show for the squeezed bath.

Additionally to (\ref{constraint}), we assume a nondegenerate representation for $\rho^*$ in terms of a basis $\{|p_i\rangle \}$,
\begin{equation}
\label{constraintB}
    \rho^* = \sum_i p_i |p_i\rangle \langle p_i|,
\end{equation}
such that $0<p_i<1$ and $p_i \neq p_j$, for any $i\neq j$. 
For the bound to be finite, we also assume the system satisfies 
\begin{equation}
    \label{constraintC}
  \langle \lambda |\rho_S(t) |\lambda \rangle < 1,
\end{equation}
for $\lambda \in \{\min_i p_i, \max_i p_i\}$. For infinite dimensional systems $\min_i p_i$ might not exist, then the assumption is checked only for $\lambda=\max_i p_i$.

Formally, we find a process that maximizes the entropy production (\ref{Entropy}) with constraints $\rho_S(0)$ and $\Phi$, assuming (\ref{constraint}),(\ref{constraintB} and (\ref{constraintC}). For that purpose, we consider a process $\tilde{\rho}(t)$ starting at $\tilde{\rho}(0)=\rho(0)$, for $t=0$, and following the ansatz $\tilde{\rho}(t)=\sigma(t)$ for $t>0$, with  $\sigma(t)$ given by
\begin{equation}
\label{solution}
    \sigma(t) :=  \frac{\exp(\alpha(t) \ln \rho^*)}{Z(\alpha(t))}=\sum_i \frac{p_i^{\alpha(t)}}{Z(\alpha(t))} |p_i\rangle \langle p_i|,
\end{equation}
where $Z(\alpha)=\tr(\exp(\alpha (\ln \rho^*)))$ and the real parameter $\alpha(t)$ is defined using (\ref{constraint}), 
\begin{equation}
\label{constraint2}
    \tr\{\rho_S(t) \ln \rho^*\}=\tr\{\sigma(t) \ln \rho^*\}=\frac{d}{d \alpha }\ln Z(\alpha):=f(\alpha).
\end{equation}
In order to show the process $\tilde{\rho}(t)$ is well defined, we prove the existence and uniqueness of $\alpha(t)$ below. Then, we show $\tilde{\rho}(t)$ maximizes the entropy production (for the same flux $\Phi$) and find the underlying upper bound. 

{\bf \emph{Existence and uniqueness -}}
For the existence of $\alpha(t)$, we write $f(\alpha)$ from (\ref{constraint2}) explicitly
\begin{equation}
f(\alpha)=\sum_i \frac{p_i^\alpha}{Z(\alpha)}\ln p_i =\sum_i \tilde{p}_i \ln p_i,
\end{equation}
for $\tilde{p}_i:=p_i^\alpha/Z(\alpha)$, $f:\mathbb{R}\rightarrow Im(f)$ has limits $\lim_{\alpha \rightarrow \infty} f(\alpha) =\ln \max_i p_i$ and $\lim_{\alpha \rightarrow -\infty} f(\alpha) =\ln \min_i p_i$, as $0<p_i<1$ from (\ref{constraintB}). Since $f$ is continuous, we have $(\ln \min_i p_i; \ln \max_i p_i)\subset Im(f)$. 
Finally, we rewrite the l.h.s. from (\ref{constraint2}) explicitly, defining $q_i:=\langle p_i | \rho_S(t)|p_i\rangle$,
\begin{equation}
\label{boundf}
\ln \min_i p_i < \sum_i q_i \ln p_i < \ln \max_i p_i,
\end{equation}
where the bounds follow from $\sum_i q_i = 1$, $0 \leq q_i \leq 1$ and the assumption (\ref{constraintC}). For infinite dimensional systems, $\min_i p_i$ might not be defined, and the lower bound in (\ref{boundf}) is $-\infty$ without loss of generality. As $f(\alpha)$ is continuous and $\tr\{\rho_S(t)\ln \rho^*\} \in (\ln \min_i p_i; \ln \max_i p_i) \subset Im(f)$ from (\ref{boundf}), there exists a $\alpha \in \mathbb{R}$ that solves (\ref{constraint2}), $f(\alpha)=\tr\{\rho_S(t)\log \rho^*\}$.

For the uniqueness of $\alpha$, we find the derivative $f'(\alpha)$,
\begin{equation}
    f'(\alpha)=\frac{\partial^2}{\partial^2 \alpha} \ln Z(\alpha) = \sum_i \tilde{p}_i (\ln p_i)^2 - (\sum_i \tilde{p}_i \ln p_i)^2 > 0,
\end{equation}
where the last inequality follows from assumption (\ref{constraintB}), which one might recognize as the variance $\langle \ln p_i^2 \rangle - \langle \ln p_i \rangle^2$, with averages taken over $\tilde{p}_i$. Since $f'(\alpha)>0$, actually $(\ln \min_i p_i; \ln \max_i p_i) = Im(f)$ and the inverse value theorem guarantees that $f(\alpha)=\tr\{\rho_S \log \rho^*\}$ for $\tr(\rho_S \log \rho^*) \in Im(f)$ has a unique solution.

On a side note, the particular cases for which $f'(\alpha)=0$ are found when all the eingenvalues of $\rho^*$ are equal (uniform distribution) or $\rho^*$ is a pure state, meaning $\alpha(t)$ could not be defined. Both degenerate situations are better understood when $\rho^*$ is thermal: they are equivalent to infinite and zero temperature baths, respectively. For that reason, they were excluded in the assumption (\ref{constraintB}).

{\bf \emph{Upper bound-}}
Now that we proved the existence and uniqueness of the process $\tilde{\rho}(t)$, we show it maximizes the entropy production. First, using (\ref{Entropy}) and (\ref{constraint}), we write the entropy production for the process $\tilde{\rho}(t)$,
\begin{equation}
\label{integralproduction0}
    \tilde{\Sigma}= -\alpha(t) \tr (\sigma(t)\ln \rho^*) + \ln Z(\alpha(t)) - S(\rho_S(0)) + \Phi.
\end{equation}
Now consider the relative entropy $D(\rho||\sigma)=\tr \rho \ln \rho - \tr \rho \ln \sigma$,
\begin{equation}
\label{relentropy}
    D(\rho_S(t)||\tilde{\rho}(t))=-S(\rho_S(t))+\ln Z(\alpha(t))- \alpha(t) \tr(\rho_S(t)\ln \rho^*).
\end{equation}
Rearranging terms in (\ref{relentropy}) and using (\ref{integralproduction0}) and (\ref{constraint}) one obtains
\begin{equation}
\label{upperbound00}
    D(\rho_S(t)||\tilde{\rho}(t))= \tilde{\Sigma}-\Sigma \geq 0,
\end{equation}
 as $D(\rho||\sigma) \geq 0 $ for any $\rho, \sigma$. Thus, Eq. (\ref{upperbound00}) proves the upper bound, $\Sigma \leq \tilde{\Sigma}$.
 The bound can be explicitly written in terms of the entropy flux, the initial state $\rho_S(0)$, and $\rho^*$,
 \begin{equation}
 \label{integralproduction01}
    \tilde{\Sigma}= \Phi-\alpha(t)\Phi- \alpha(t)\tr(\rho_S(0)\ln\rho^*)+\ln(Z(\alpha(t))-S(\rho_S(0)).
\end{equation}
Analogously, the bound is also written in terms of the entropy flux rate $\phi_i(t)=d\Phi_i/dt$,
 \begin{equation}
\label{integralproduction2}
    \tilde{\Sigma}=\Sigma_a+ \int_0^t (1-\alpha(t))\phi(t)dt,
\end{equation}
where $\Sigma_a:=S[\sigma(0)]-S[\rho_S(0)]\geq 0$. 

For the specific case of thermal environments, a similar expression appears in the context of finite baths \cite{Strasberg2021A,Strasberg2021B,Strasberg2021C} from the Clausius inequality, where the temperature of the bath is time-dependent. Equation (\ref{integralproduction2}) is the general form of the upper bound (\ref{upperbound}), and it consists of our main result: the process $\tilde{\rho}(t)$ maximizes the entropy production with respect to the constraint (\ref{constraint2}). Moreover, as $D(\rho_S(t)||\tilde{\rho}(t))=0$ if and only if $\rho_S(t)=\tilde{\rho}(t)$, we have the upper bound attained only for the maximal process. 

An alternative version of the result considers, instead of (\ref{constraint}), the entropy flux rate as a constrained observable with the form
\begin{eqnarray}
\label{constraint2rate}
\phi:=\frac{d\Phi}{dt}=\tr(\rho_S (t) \ln \rho^*)+C,
\end{eqnarray}
for some constant $C$, resulting in the same constraint (\ref{constraint2}) upon matching the entropy flux rates of $\rho_S(t)$ and $\sigma(t)$, without loss of generality. In this case, the resulting bound $\tilde{\Sigma}$ (\ref{integralproduction2}) (see Appendix A) is given as
\begin{equation}
\label{integralproduction3}
    \tilde{\Sigma}=\Sigma_a+ \int_0^t [\phi(t)-\alpha(t)\dot{\phi}(t)]dt.
\end{equation}
This last representation (\ref{integralproduction3}) is particularly useful for the squeezed reservoir, as discussed later in the paper.

\begin{figure}[htp]
\includegraphics[width=3.3 in]{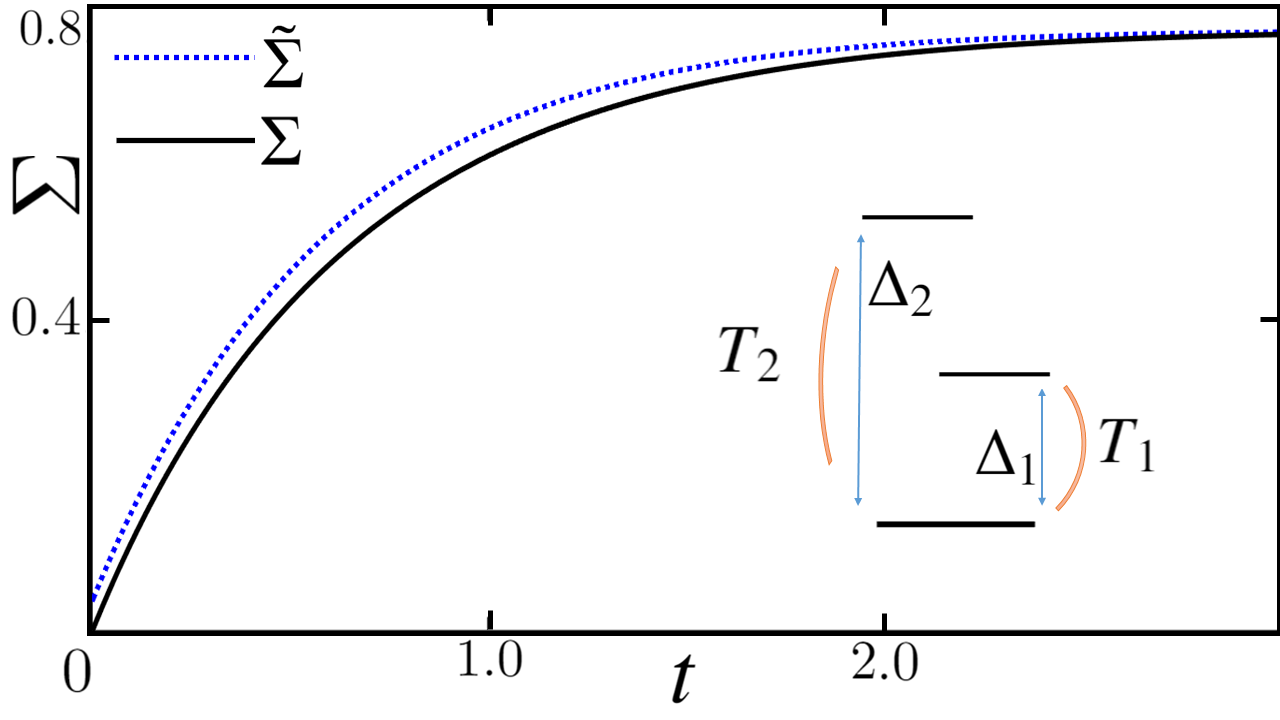}
\caption{(Color online) The entropy production of a three level maser as a function of time from a initial state with coherence $(p_1(0),p_2(0),c(0))=(0.3,0.3,0.1)$ and energy gaps $\Delta_1=1$, $\Delta_2=2$. The system is weakly coupled to two reservoirs at temperatures $T_1=1$ and $T_2=0.5$ with parameters in the dynamics $\lambda_1=\lambda_2=1$, $\Gamma_d=1.2$. The gap between the upper bound and the actual entropy production tends rapidly to zero over time due to decoherence.}
\label{Fig2d.png}
\end{figure}
{\bf \emph{Applications -}} 
In the following examples, we check that the entropy flux satisfies (\ref{constraint}) for some $\rho^*$ satisfying (\ref{constraintB}), then we compute the upper bound (\ref{integralproduction2}) with $\alpha(t)$ defined from (\ref{constraint2}). We will use the shorthand notation $\rho(t)$ to represent the system's density matrix, $\rho_S(t)$.

\textit{1. Single qubit--} Consider the single qubit with energies $E_0=-\Delta/2$ and $E_1=\Delta/2$ weakly coupled to a thermal environment. The system evolves with the following density matrix:
\begin{equation}
\label{qubitmatrix}
    \rho(t) = \begin{pmatrix} 
    1-p(t) & c(t) \\[0.2cm] c(t)^* & p(t) \end{pmatrix}, 
\end{equation}
in the Hamiltonian eigenbasis. The entropy flux is given by $\Phi=\beta Q_E$. Due to weak coupling approximation, we have $\Phi=-\beta \Delta (p(t)-p(0))$, which has the form (\ref{constraint}), for $\rho^*=C[|0\rangle\langle0|+\exp(-\beta \Delta)|1\rangle \langle 1|]$ and $C=(1+\exp(-\beta \Delta))^{-1}$. Therefore, the system satisfies the requirement for the upper bound.
The master equation dynamics reads $p(t)=(p_0-p_\infty)e^{-\gamma t}+p_{\infty}$ and $c(t)=c(0) e^{-i\Delta t-\gamma t/2}$, where $\{1-p_0,p_0\}$ are the initial populations. From (\ref{integralproduction2}), the upper bound reads
\begin{equation}
\label{upperboundqubit}
    \tilde{\Sigma} = \Sigma_a  +\gamma \beta\Delta (p_0-p_\infty)\int_0^t (1-\alpha(t)) e^{-\gamma t}dt,
\end{equation}
where we used $\phi(t)= \gamma \Delta (p_0-p_\infty)e^{-\gamma t} $ and $\alpha(t)$ given from (\ref{constraint2}), which results in
\begin{equation}
 \tanh\big(\frac{\alpha(t)\beta\Delta}{2}\big)=1-2p(t),
\end{equation}
and the constant term $\Sigma_a$ given by
\begin{equation}
\Sigma_a=- p(0)\ln p(0) -(1-p(0))\ln (1-p(0)) +\lambda_+ \ln \lambda_+ +\lambda_- \ln \lambda_-,
\end{equation}
with $\lambda_{\pm}=\frac{1}{2}\pm \sqrt{(p(0)-\frac{1}{2})^2+|c(0)|^2}$, which is the initial relative cost of coherence. Note that the bound (\ref{upperboundqubit}) is given solely in terms of $\rho(0)$ and $p_\infty$ and $\Phi$, since $\Phi$ defines $p(t)$ and $\alpha(t)$ uniquely,

\textit{2. Three-level amplifier-} Consider the three level amplifier \cite{Scovil1958,Geusic1967,Mitchison2019,Li2017a}, consisting of three levels $\{|0\rangle, |1\rangle, |2\rangle \}$. The levels ($0,1$) are weakly coupled to a cold reservoir ($T_c$), levels ($0,2$) are weakly coupled to a hot reservoir ($T_h$) and levels ($1,2$) are coupled to an external field. We are interested in the relaxation process where the field is suddenly turned off from any initial condition. In the relaxation, the Hamiltonian is given by $H=H_0=E_0|0\rangle \langle 0\rangle + E_1|1\rangle \langle 1|+E_2|2\rangle \langle 2|$. The nonequilibrium density matrix $\rho(t)$ reads for $t \geq 0$,
\begin{equation}
\label{3matrix}
    \rho(t) = \begin{pmatrix} 
    p_0(t) & 0 & 0 \\[0.2cm] 
    0 & p_1(t) & c(t) \\[0.2cm] 
    0 & c(t)^* & p_2(t) \\[0.2cm] 
    \end{pmatrix}, 
\end{equation}
which is the solution of a quantum master equation \cite{Kosloff2014,Kosloff1996},
\begin{eqnarray} 
\label{master}
\dot{p_1}=-\lambda_1 (1+e^{-\beta_1 \Delta_1})p_1+\lambda_1 e^{-\beta_1 \Delta_1}(1-p_2),
\\ \dot{p_2}=-\lambda_2 (1+e^{-\beta_2 \Delta_2})p_2+\lambda_2 e^{-\beta_2 \Delta_2}(1-p_1),
\\ \dot{c}=i(\Delta_1-\Delta_2) c -\Gamma_d c,
\end{eqnarray}
for constants $\lambda_1,\lambda_2,\Gamma_d$, $\Delta_i=E_i-E_0$ and we assumed the same damping coefficient $\gamma$ for both levels, $n_i=1/(\exp(\beta_i\Delta_i)+1)$. The entropy flux given by $\Phi=\beta_1 Q_1+\beta_2 Q_2$. Due to weak coupling approximation, we have $\Phi=-\beta_1 \Delta_1 (p_1(t)-p_1(0))-\beta_2 \Delta_2 (p_2(t)-p_2(0))$, which satisfies (\ref{constraint}) for $\rho^*=C[|0\rangle \langle 0|+\exp(-\beta_1\Delta_1)|1\rangle\langle1| +\exp(- \beta_2\Delta_2)|2\rangle\langle 2|)]$, $C=[1+\exp(-\beta_1\Delta_1)+\exp(-\beta_2\Delta_2)]^{-1}$, checking the requirement for the bound (\ref{integralproduction2}) with a single constraint, $\Phi$. Now we have to find $\Sigma_a$ and $\alpha(t)$.

First, the state $\sigma(0)$ is defined in (\ref{solution}) with $\alpha_0$ given from the solution of the transcendental equation
\begin{equation}
\label{findingalpha}
   p_1(0)\beta_1\Delta_1+p_2(0)\beta_2\Delta_2=\frac{\beta_1\Delta_1e^{-\alpha_0 \Delta_1\beta_1}+\beta_2\Delta_2e^{-\alpha_0 \Delta_2\beta_2}}{1+e^{-\alpha_0\beta_1\Delta_1}+e^{-\alpha_0\beta_2\Delta_2}},
\end{equation}
where the lhs is a constant as it depends on the initial state. We use $\alpha_0$ to find the constant term $\Sigma_a=S(\sigma(0))-S(\rho_0)$. For $t>0$, for computational purposes, instead of solving a transcendental equation like (\ref{findingalpha}), it is useful to take $\alpha_0$ and compute $\alpha(t)$ and the bound incrementally for $t>0$. The increment of entropy flux is taken from (\ref{constraint})
\begin{equation}
    \delta \Phi = \tr\{(\rho(t+dt)-\rho(t))\ln \rho^*\},
\end{equation}
where $\rho(t)$ comes from the master equation (\ref{master}) and the flux is updated as $\Phi(t+dt)=\Phi(t)+\delta \Phi$. The update in $\alpha(t)$ is easily written in terms of the flux increment from (\ref{constraint2}) and the bound increment obtained from (\ref{integralproduction2})
\begin{eqnarray} 
 \alpha(t+dt)=\alpha(t)+ \delta \Phi/g(\alpha),
\\
\tilde{\Sigma}(t+dt)=\tilde{\Sigma}(t)+(1-\alpha(t))\delta \Phi,
\end{eqnarray}
where $g(\alpha)=\partial^2_\alpha \ln Z(\alpha)$ and $\tilde{\Sigma}(0)=\Sigma_a$. In Fig.~1, we plot the bound $\tilde{\Sigma}(t)$ vs. the actual entropy production $\Sigma(t)$ computed from definition (\ref{Entropy}) for different times. The algorithm above shows a nonpolynomial relation between the rate of the entropy production bound and the underlying entropy flux rate, in the same spirit of the relation obtained for Gaussian systems \cite{SalazarLandi2020}. The tightness $\tilde{\Sigma}-\Sigma$ at $t=0$ is given by the term $\Sigma_a$, and it decreases with time until it vanishes for long times.

\begin{figure}[htp]
\includegraphics[width=3.3 in]{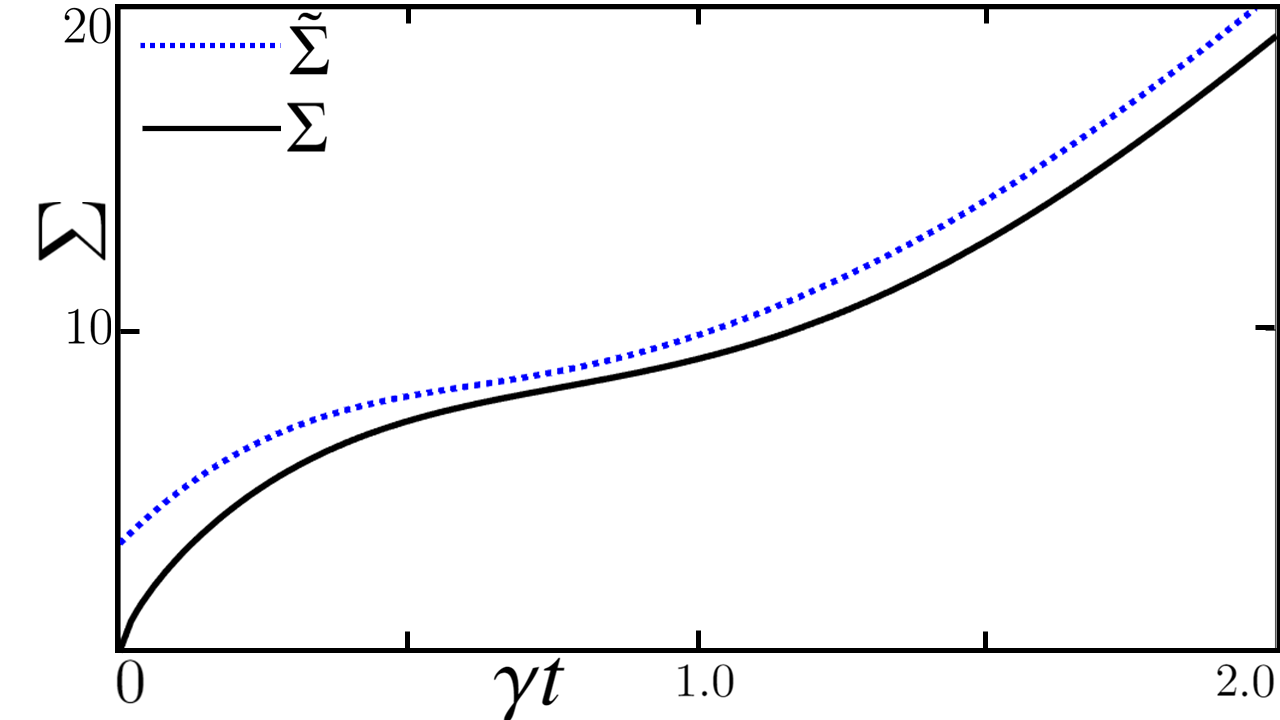}
\caption{(Color online) The entropy production $\Sigma$ as a function of time for a bosonic mode ($\hbar\omega/k_bT=2)$ coupled to a nonequilibrium broadband squeezed bath ($\hbar\omega_s/k_bT=1$ and $r=1, \theta=1$) for initial conditions $\rho_0=S(\epsilon)|0\rangle \langle 0| S^\dagger(\epsilon)$, $\epsilon=-r$. The gap between the upper bound $\tilde{\Sigma}$ and the actual entropy production $\Sigma$ remains finite for long times as a signature of the nonequilibrium steady state.}
\end{figure}

\textit{3. Squeezed bath-} Now we apply the result to a bosonic mode with Hamiltonian $H=\omega(a^\dagger a +1/2)$ in contact with a nonequilibrium broadband squeezed bath, satisfying the Lindblad's dynamics,
\begin{equation}
\label{Lindblads}
\frac{d \rho}{dt}=-i[H,\rho]+ D(\rho),
\end{equation}
where the dissipator $D(\rho)$ is defined for a squeezed bath
\begin{equation}
\label{dissipator}
D(\rho)=\gamma(\overline{n}+1)\big[b_z \rho b_z^\dagger-\frac{1}{2}\{b_z^\dagger b_z,\rho\}\big]+\gamma \overline{n}\big[b_z^\dagger \rho b_z - \frac{1}{2}\{b_z b_z^\dagger, \rho\}\big],
\end{equation}

where $\overline{n}=(e^{\omega/T}-1)^{-1}$ is the bosonic thermal occupation number for the temperature $T$ and $b_z=S(z)aS(z)^\dagger$, for $S(z)=\exp\{(z^*a^2-za^{\dagger 2})/2\}$ and $z=r\exp(i(\theta-2\omega_s t))$, where $r\exp(i\theta)$ is the squeezed parameter of the bath with central frequency $\omega_s$.

We are interested in Gaussian initial conditions, which makes the dynamics (\ref{Lindblads}) particularly simple, because it maps Gaussian states into Gaussian states for $t>0$. In this case, the state for $t>0$ uniquely defined in terms of first and second order observables \cite{Santos2017a},
\begin{eqnarray} 
\label{squeezedME}
\frac{d}{dt}\langle a\rangle = -(\frac{\gamma}{2}+i\omega)\langle a \rangle,
\\
\frac{d}{dt}\langle a^\dagger a\rangle = -\gamma (\langle a^\dagger a \rangle -N),
\\
\frac{d}{dt}\langle a a\rangle = -(\gamma +2i\omega) \langle aa \rangle + M_0 e^{-2i\omega_s t},
\end{eqnarray}

with $M_t=-(\overline{n}+1/2)\exp(i(\theta-2\omega_s t))\sinh(2r)$ and $N+1/2=(\overline{n}+1/2)\cosh(2r)$.
The entropy flux is given by $\Phi=\beta\omega \gamma\int_0^t (\langle b_z^\dagger b_z\rangle_t - \overline{n})dt$, which satisfies $\phi=\tr \{\rho(t)\ln \rho^*\}-C$ for $\ln \rho^*=\gamma \omega \beta (b_z^\dagger b_z - \overline{n}I)+C$, for some normalization constant $C$, checking the requirement for the upper bound (\ref{constraint2rate}). From (\ref{integralproduction3}), the upper bound reads
\begin{equation}
\label{upperboundsqueezed}
    \tilde{\Sigma} = \Sigma_a  +\Phi -\int_0^t \alpha(t)\dot{\phi}dt,
\end{equation}
in terms of flux related quantities and initial conditions. The state $\sigma(t)$ given in (\ref{solution}) is thermal (in the basis $b_z$) with $\alpha(t)$ given by (\ref{constraint2}), resulting in
\begin{equation}
\label{alphasqueezed}
\alpha(t) \beta \omega \gamma = -\ln \Big(\frac{1+\langle b_z^\dagger b_z\rangle_t}{\langle b_z^\dagger b_z\rangle_t}\Big).
\end{equation}
Note that, as $\phi=\gamma \omega \beta(\langle b_z^\dagger b_z \rangle - \overline{n})$, $\alpha(t)$ is a function of the entropy flux rate in (\ref{alphasqueezed}). Using the entropy for a thermal state, we obtain from the initial condition $\Sigma_a = S(\sigma(0))-S(\rho(0))$. Considering the case where $\rho(0)$ is a pure state, $\Sigma_a=S(\sigma(0))$, we get  
\begin{equation}
\label{sigmaasqueezed}
\Sigma_a=(\langle b_z^\dagger b_z\rangle_0+1)\ln(\langle b_z^\dagger b_z\rangle_0+1)-\langle b_z^\dagger b_z\rangle_0\ln \langle b_z^\dagger b_z\rangle_0.
\end{equation}
Combining $\Sigma_a$ from (\ref{sigmaasqueezed}) and $\alpha(t)$ from (\ref{alphasqueezed}), the bound (\ref{upperboundsqueezed}) is written in terms of the initial condition and the entropy flux rate (or equivalently, in terms of the average occupation number $\langle b^\dagger b \rangle_t=\langle a^\dagger a\rangle_t \cosh(2r)+\sinh(r)^2 -Re[M(t)^*\langle a a\rangle_t/(\overline{n}+1/2)]$). In Fig.~2, we show the entropy production (\ref{upperboundsqueezed}) as a function of time for the system starting from a pure squeezed state, $\rho(0) =S(\epsilon)|0\rangle \langle 0|S(\epsilon)^\dagger$, $\epsilon=-r$. For this initial condition, we have $\langle a \rangle_0 = 0$, $\langle a^\dagger a \rangle_0=\sinh(r)^2$, $\langle a a \rangle_0=\sinh(r)\cosh(r)$. The actual entropy production $\Sigma(t)$ is computed in the Appendix B.

{\bf \emph{Discussion -}} 
In this paper, we derived an upper bound for the entropy production written in terms of the entropy flux (or entropy flux rate), and the initial state for a class of systems for which the entropy flux is written as a system's observable. The upper bound is closely related to the MaxEnt principle, and, for thermal reservoirs, the bound is particularly attained for time-dependent thermal states. As a matter of fact, it is interesting that the notion of time-dependent ``temperature" $\alpha(t)$ is present even in a nonthermal environment. We expect the bound to be useful from an experimental perspective, as the entropy fluxes are easier to assess than the actual entropy production.

{\bf \emph{Acknowledgements -}} 
The author thanks G. T. Landi for fruitful discussions.

{\bf \emph{Appendix A -}}
The alternative version of the bound considers the entropy flux $d\Phi/dt:=\phi(t)=\tr (\rho_S(t) \ln \rho^*)+C$ as constraint (\ref{constraint2rate}). Upon matching this flux rate to the flux rate of the maximal process yields $\tr (\rho_S(t) \ln \rho^*)=\tr(\sigma(t)\ln \rho^*)$, which is the same constraint as (\ref{constraint2}) and results in a $\alpha(t)$. A useful representation of the bound was given in (\ref{integralproduction3}) and the proof goes as follows. Take the definition (\ref{Entropy}) for the process $\tilde{\rho}(t)$,
\begin{equation}
\label{tildesigmaA}
    \tilde{\Sigma} = \Sigma_a + \Phi + \int_0^t \frac{dS}{dt}dt,
\end{equation}
where we used $\Delta S=\int_0^t (dS/dt) dt
= -\int_0^t \tr(\dot{\sigma} \ln \sigma)dt$. Now using the definition of $\sigma(t)$ from (\ref{solution}), we obtain
\begin{equation}
    \tr(\dot{\sigma}\ln \sigma)=\alpha(t)\tr(\dot{\sigma}\ln \rho^*)=\alpha(t)\tr(\dot{\rho}\ln \rho^*)=\alpha(t)\dot{\phi},
\end{equation}
where the last identity follows from the constraint (\ref{constraint2}). Then, the upper bound (\ref{tildesigmaA}) now reads.
\begin{equation}
\label{tildesigmaB}
    \tilde{\Sigma} = \Sigma_a + \Phi - \int_0^t \alpha(t)\dot{\phi}(t)=\Sigma_a + \int_0^t(\phi(t)-\alpha(t) \dot{\phi}(t))dt ,
\end{equation}
which is the representation (\ref{integralproduction3}), using $\Phi=\int_0^t \phi(t)dt$.

{\bf \emph{Appendix B -}} The upper bound for the squeezed bath was computed in the main text. The actual entropy production $\Sigma$ requires some steps. First, note that $S(\rho(0))=0$ from initial conditions. The entropy production is given by
\begin{equation}
\label{actualproduction}
    \Sigma = S(\rho(t))+\Phi.
\end{equation}
$\rho(t)$ is a gaussian state with a general form
\begin{equation}
  \rho(t)=C \exp(-\lambda(ca^\dagger +d a)(c^*a+d^*a^\dagger)),
\end{equation}
without displacement terms (first order in $a$, $a^\dagger$, from our initial conditions), where we impose $|c|^2-|d|^2=1$. This form is particularly useful for the diagonalization, since $B:=ca^\dagger+da$ satisfies the commutation relation $[B,B^\dagger]=1$. Therefore, the density matrix is written in a diagonal form in the eigenbasis of $B^\dagger B$,
\begin{equation}
  \rho(t)=\frac{\sum_0^\infty e^{-\lambda n} | n_B \rangle \langle n_B |}{\sum_0^\infty e^{-\lambda n}},
\end{equation}
such that $\langle n_B |B^\dagger B |n_B \rangle = n_B$.
Now we obtain $\lambda$ from the observables of our problem $\langle a^\dagger a \rangle$, $\langle a a \rangle$ using the inverse transformations
\begin{eqnarray}
a^\dagger = c^*B^\dagger - dB\\
a=-d^*B^\dagger + cB,
\end{eqnarray}
which results in
\begin{eqnarray}
\langle a^\dagger a\rangle = |d|^2+(1+2|d|^2)u,\\
|\langle a a \rangle|^2 = |d|^2|c|^2(1+2u)^2,
\end{eqnarray}
where we defined $u=\langle B^\dagger B\rangle=e^{-\lambda}/(1+e^{-\lambda})$. It results in a polynomial equation for $u$,
\begin{eqnarray}
(\langle a^\dagger a\rangle - u)(\langle a^\dagger a\rangle +u-1)=|\langle a a \rangle|^2
\end{eqnarray}
Now that $u$ is given in terms of the observables $\langle a^\dagger a\rangle$ and $\langle a a \rangle$, the entropy is given by
\begin{eqnarray}
S(\rho(t))=(1+u)\ln(1+u)-u\ln u.
\end{eqnarray}
Inserting the entropy in (\ref{actualproduction}) together with  the flux $\Phi$ (constraint), we get the actual entropy production for the process.
\bibliography{library}

\end{document}